\documentclass[conference,10pt,  onecolumn]{IEEEtran}

 \usepackage[pdftex]{graphicx}
 \usepackage[cmex10]{amsmath}
 \usepackage{amssymb}
 \usepackage{xcolor}
\usepackage{amsthm}
\usepackage{algorithmic}
\usepackage{mathrsfs}

\allowdisplaybreaks[4]

\theoremstyle{plain}

\theoremstyle{definition}

\theoremstyle{remark}

\usepackage{parskip}

\title{Source-channel separation for two-way interactive communication with fidelity criteria}

 \author{Mukul Agarwal}

\begin{document}

\maketitle

\begin{abstract}

Consider the channel coding problem where two users are interacting in order to communicate an i.i.d. source $X_1$ from User 1 to User 2 with distortion $D_1$ and an i.i.d. source $X_2$ from from User 2 to User 1 with distortion $D_2$. $X_1$ and $X_2$ may be dependent.  Communication occurs from User 1 to User 2 via a memoryless channel $\mathcal C_1$ and from User 2 to User 1 over a memoryless channel $\mathcal C_2$, where $\mathcal C_1$ and $\mathcal C_2$ are independent of each other. Communication occurs during each time slot between both users and each user can make a codin and decoding based on all past available knowledge. This interactive communication problem is formulated and it is proved that source-channel separation based architectures are optimal.

\end{abstract}

\section{Introduction}

We consider the channel coding problem where two users are interacting in order to communicate an i.i.d. source $X_1$ from User 1 to User 2 with distortion $D_1$ and  an i.i.d. source $X_2$ from User 2 to User 1 with distortion $D_2$ where the sources $X_1$ and $X_2$ may be dependent, and the communication from User 1 to User 2 occurs over a memoryless channel $\mathcal C_1$, the communication from User 2 to User 1 occurs over a memoryless channel $\mathcal C_2$, where the channels $\mathcal C_1$ and $\mathcal C_2$ are independent of each other. A precise mathematical formulation for this problem of communication is provided in Section \ref{Formulation}. In Sections \ref{Staggered} and \ref{SCS}, it is proved that we can restrict attention to a certain class of codes that we called staggered codes, wherein, if $\mathcal C_1$ is being used for communication from User 1 to User 2, the $\mathcal C_2$ is not being used for communication from User 2 to User 1, and similarly, if $\mathcal C_2$ is being used for communication from User 2 to User 1, then $\mathcal C_1$ is not being used for communication from User 1 to User 2. The source-coding problem corresponding to this form of communication where User 1 and User 2 do not both act at the same time has been studied in \cite{Kaspi}; see also \cite{KimGamal}, Pages 514-519. We use converse-style mutual-information arguments and these source-coding results in the manner described in \cite{KimGamal}, Pages 514-519, in order to prove that the required interactive  communication can be carried out in a manner of source-channel separation without loss of optimality, and this is the subject of Section \ref{SCS}.

Some of the information and coding literature on 2-way communication is the following. Two-way commuication was considered in \cite{2Way}, where an infinite-letter characterization was provided for its capacity region. The problem of 2-way source-coding with fidelity criteria was considered, as stated above, in \cite{Kaspi}, where the $q$-round rate-distortion region for two-way lossy source-coding was established. Interactive function computation is considered in \cite{MaIshwar}, wherein a `computable' characterization of the rate-region is provided. The problem of coding for interactive communication has been considered in \cite{Schulman}, and the reader is referred to \cite{Braverman} for definitions and results concerning complexity and two-way interactive communication from a Computer Science perspective.

Much work exists in the information theory literature on source-channel separation, both for lossless and lossy communication; we restrict here to giving references for the case of lossy communication:  The seminal work is \cite{Shannon} where, source-channel separation theorem is proved for the point-to-point DMC for communication of an i.i.d. source. For the general network, source channel separation has been proved to hold in the unicast setting for memoryless networks, see \cite{T} and for general, unknown networks, see \cite{A1}. By unicast setting (called multiple-unicast in \cite{T}), is meant, the scenario that sources$X_{ij}$, where $X_{ij}$ denotes the source which needs to be communicated from User $i$ to User $j$, are independent of each other. When the sources are dependent, source-channel separation is known to not hold in general, even in the lossless setting, see for example \cite{Gastpar}. In this paper, we show a case where even if the sources are dependent, source-channel separation holds, and that is the problem of two-way communication with fidelity criteria.

There are various other references, possibly numerous, especially on source-channel separation, for which, the authors seek pardon for not referencing.

\section{The mathematical formulation of the problem of $2$-way  interactive communication with fidelity criteria} \label{Formulation}

There are two users User 1 and User 2. $\mathcal{C}_1$ and $\mathcal{C}_2$ are the two discrete memoryless channels. $\mathcal C_1$ can be used for communication from User 1 to User 2. $\mathcal C_2$ can be used for communication from User 2 to User 1. The Shannon capacities of $\mathcal C_1$ and $\mathcal C_2$ are $C_1$ and $C_2$ respectively. The input space and output spaces of $\mathcal C_1$ are $\mathbb I_1$ and $\mathbb O_1$ respectively. The input and output spaces of $\mathcal C_2$ are $\mathbb I_2$ and $\mathbb O_2$. $\mathbb I_1, \mathbb O_1, \mathbb I_2, \mathbb O_2$ are assumed to be finite sets.  For $x  \in \mathbb I_1$, the action of the channel $\mathcal C_1$ on $x$ results in an output denoted by $\mathcal C_1x$ or $\mathcal C_1(x)$. Thus,  $\mathcal C_1x \in \mathbb O_1$. For $x^n \in \mathbb I_1^n$, the action of $\mathcal C_1$ on $x^n$ is denoted by $\mathcal C_1( x^n)$, which is the same as $(\mathcal C_1(x_1), \mathcal C_1(x_2), \ldots, \mathcal C_1(x_n))$. A similar notation is used for action of $\mathcal C_2$. User $1$ has an i.i.d. source input $X_1$ and user $2$ has an i.i.d. source input $X_2$. $X_1$ and $X_2$ may be dependent. When the block length is $n$, the i.i.d. sources are $X_1^n$ and $X_2^n$ respectively. User $1$ needs to communicate the i.i.d. $X_1$ source to User 2 and User 2 needs to communicate i.i.d. $X_2$ source to User 1 by communication over channels $C_1$ and $C_2$ in an interactive manner which will become clearer in the definitions below.

The definition below has been written in a repetitive manner for the sake of clarity.

For block-length $n$, a channel-code for $2$ way interactive communication consists of$\{N, <f_{1i}>_1^N, <f_{2i}>_1^N, c_{1i}>_1^N, <c_{2i}>_1^N, g_1, g_2\}$, where this notation means the following: $N$ is the time horizon. $c_{ij}$ is either $1$ or  $0$; i = {1, 2}, $j \in \{1, 2, \ldots, N \}$ and at least one of $c_{1i}$ and $c_{2i}$ is $1$, $1 \leq i \leq N$. If $c_{1j} = 1$,  communication occurs from User 1 to User 2 during time slot $j$, else it does not occur. Similarly, if $c_{2j} = 1$, communication occurs from User 2 to User 2 during time slot $j$, else it does not occur.  $f_{1i}$ and $f_{2i}$ are the functions which cause 2-way interactive communication;  $g_1, g_2$ are the source reproduction functions. The knowledge of functional forms of $f_{1i}, f_{2i}, g_1, g_2$, the values of $N$, $c_{1i}, c_{2i}$, $1 \leq i \leq N$, are assumed to be known apriori  at User 1 and User 2, and will not be shown explicitly in what follows.

$f_{1i}$, $1 \leq i \leq N$, denotes the function (encoder) which is used for mapping all information available at User 1 until time $i$  into an input for the channel $\mathcal C_1$ at time $i$. Similarly, $f_{2i}$, $1 \leq i \leq N$, denotes the function (encoder) which is used for mapping all information available at User 2 until time $i$ into an input for the channel $\mathcal C_2$ at time $i$. The function (decoder)$g_1$ produces an estimate of the  $n$-length source $X_2$ at User 1 and function (decoder) $g_2$ produces an estimate of the $n$-length source $X_1$ at User 2. Mathematically stated definition of $f_{1i}, f_{2i}, g_1, g_2$ follow.

If $c_{11} = 1$, $u_{11} = f_{11}(x_1^n)$, denote $v_{11} = \mathcal C_1(u_{11})$;  else, $u_{11} = e$ and $v_{11} = e$.

If $c_{21} = 1$, $u_{21} = f_{21}(x_2^n)$, denote $v_{21} = \mathcal C_2(u_{21})$;  else, $u_{21} = e$, $v_{21} = e$.

The above two lines should be interpreted as follows: During time slot $1$, if $c_{11} = 1$, an encoder $f_{11}$ maps all available knowledge at User 1 (only the source input, $x_1^n$ so far) into the channel input, denoted by $u_{11}$. The channel $\mathcal C_1$ acts on this input and produces an output $v_{11}$ which is available at User 2. If $c_{11}$ is $0$, there is no input to $\mathcal C_1$ in time slot $1$ and there is no output of Channel $\mathcal C_1$.  This is denoted by $e$, which should be thought of as `idle'. Another way of thinking about this is that if $c_{11} = 0$, the channel $\mathcal C_1$ is being used for other purposes, and not this $2$-way interactive communication.  Similarly, if $c_{21}$ is $1$, an encoder $f_{21}$ maps all available knowledge at User 2 (only the source input $x_2^n$ so far) into the channel input, denoted by $u_{21}$, which is passed over the channel $\mathcal C_2$ and produces an output $v_{21}$ which is available at User 1. If $c_{21}$ is $0$, input to $\mathcal C_2$ and output of $\mathcal C_2$ is $e$, where $e$ needs to be interpreted as above.

If $c_{12} = 1$, $u_{12} = f_{12}(x_1^n, u_{11}, v_{21})$, denote $v_{12} = \mathcal C_1(u_{12})$;  else, $u_{12} = e$, $v_{12} = e$.

If $c_{22} = 1$, $u_{22} = f_{22}(x_2^n, u_{21}, v_{11})$, denote $v_{22} = \mathcal C_2(u_{22})$;  else, $u_{22} = e$, $v_{22} = e$.

The above two lines should be interpreted as follows: During time slot $2$, if $c_{12} = 1$, an encoder $f_{12}$ maps all available knowledge at User 1 so far ($x_1^n, u_{11}, v_{21}$) into the channel input, denoted by $u_{12}$. The channel $\mathcal C_1$ acts on this input and produces an output $v_{12}$ which is available at User 2. If $c_{12}$ is $0$, the input and output of $\mathcal C_1$ is $e$ where $e$ is to be interpreted as previously. Similarly, if $c_{22}$ is $1$, an encoder $f_{22}$ maps all available knowledge at User 2 so far ( $x_2^n, u_{21}, v_{11}$) into the channel input, denoted by $u_{21}$, which is passed over the channel $\mathcal C_2$ and produces an output $v_{21}$ which is available at User 1. If $c_{21}$ is $0$, input to $\mathcal C_2$ and output of $\mathcal C_2$ is $e$, where $e$ needs to be interpreted as above.

In general, for $i, 1 \leq i \leq N$,

If $c_{1i} = 1$, $u_{1i} = f_{1i}(x_1^n, u_{11}, v_{21}, u_{12}, v_{22}, \ldots, u_{1,i-1}, v_{2,i-1})$, denote $v_{1i} = \mathcal C_1(u_{1i})$; else, $u_{1i} = e$, $v_{1i} = e$.

If $c_{2i} = 1$, $u_{2i} = f_{2i}(x_2^n, u_{21}, v_{11}, u_{22}, v_{12}, \ldots, u_{2,i-1}, v_{2,i-1})$, denote $v_{2i} = \mathcal C_2(u_{2i})$; else, $u_{2i} = e$, $v_{2i} = e$.

The above two lines should be interpreted as follows: During time slot $i$, if $c_{1i} = 1$, an encoder $f_{1i}$ maps all available knowledge at User 1 so far ($x_1^n, u_{11}, v_{21}, u_{12}, v_{22}, \ldots, u_{1,i-1}, v_{2,i-1}$) into the channel input, denoted by $u_{1i}$. The channel $\mathcal C_1$ acts on this input and produces an output $v_{1i}$ which is available at User 2. If $c_{1i}$ is $0$, the input and output of $\mathcal C_1$ is $e$ where $e$ is to be interpreted as previously. Similarly, if $c_{2i}$ is $1$, an encoder $f_{2i}$ maps all available knowledge at User 2 so far ( $x_2^n, u_{21}, v_{11}, u_{22}, v_{12}, \ldots, u_{2,i-1}, v_{1,i-1}$) into the channel input, denoted by $u_{2i}$, which is passed over the channel $\mathcal C_2$ and produces an output $v_{2i}$ which is available at User 1. If $c_{2i}$ is $0$, input to $\mathcal C_2$ and output of $\mathcal C_2$ is $e$, where $e$ needs to be interpreted as above.

At the end of $N$ rounds of $2$-way interactive communication, an estimate of $x_1^n$ is made at User 2 based on all available knowledge at User 2 via a function (decoder) $g_2$. This estimate is denoted by $\hat x_1^n$. Similarly, an estimate of $x_2^n$ is made at User 1 based on all available knowledge at User 1 via a function (decoder) $g_1$. This estimate is denoted by $\hat x_2^n$. Mathematically,

$\hat {x}_1^n = g_2(x_2^n, u_{11}, v_{21}, u_{12}, v_{22}, \ldots, u_{1N}, v_{1N})$, $\hat {x}_2^n = g_1(x_1^n, u_{21}, v_{12}, u_{22}, v_{12}, \ldots, u_{2N}, v_{1N})$

Denote, $c_1 = \frac{1}{n} \sum_{j=1} c_{ij}$, $c_ 2 = \frac{1}{n} \sum_{j=1} c_{2j}$. $c_1$ denotes the number of channel uses of $\mathcal C_1$ per input symbol of $X_1$ and $c_2$ denotes the number of channel uses of $\mathcal C_2$ per input symbol of $X_2$. 

 When the sources are random, $X_1^n$ and $X_2^n$, the above dynamic via the functions $f_{1i}, f_{2i}, g_1, g_2$, $1 \leq i \leq N$ leads to a joint vectors $(X_1^n, \hat{ X}_1^n)$ and $(X_2^n, \hat{ X}_2^n)$. $\hat X_1^n$ is the estimate of $X_1^n$ and $\hat X_2^n$ is the  estimate of $X_2^n$. In other words, $\hat X_1^n$ is the random vector corresponding to $\hat x_1^n$ and $\hat X_2^n$ is the random vector corresponding to $\hat x_2^n$.

If there exists an $n$, and a code defined as above such that $c_1 \leq R_1$, $c_2 \leq R_2$,  $\frac{1}{n} Ed_1(X_1^n, \hat{X}_1^n) \leq D_1$, $\frac{1}{n} Ed_2(X_2^n, \hat{X}_2^n) \leq D_2$, it is said that there exists a $(\mathcal C_1, \mathcal C_2, X_1, X_2, R_1, R_2, D_1, D_2)$ code for $2$-way interactive communication.

Note that $c_1$ and $c_2$ are defined in such a way that the length of the time horizon $N$ does not matter; what matters is the number of channel uses for the certain block-length $n$. For the times $C_1$ or $C_2$ is not in use in order to cause the two way interactive communication, the channels can be used for other purposes.

\section{Staggered code} \label{Staggered}

The code defined in the previous section  is said to be staggered if  exactly one  of  $c_{1i} $ or $c_{2i} = 1$, $1 \leq i \leq N$;  the first communication happens from User 1 to User 2, that is, $c_{11} = 1$ and $c_{21} = 0$; and  the last communication occurs from User 2 to User 1, that is, $c_{2N} = 1$ and $c_{1N} = 0$.

A good way to think of a staggered code is the following: It consists of numbers $n_1, n_2, \ldots, n_q$, $q$ even. First , communication occurs from User $1$ to user $2$ for $n_1$ slots of time. Then, communication occurs from User $2$ to User $1$ for $n_2$ slots of time. Then, communication occurs from User 1 to User 2 for $n_3$ slots of time. Then, communication occurs from User 2 to User 1 for $n_4$ slots of time. In the last two rounds, communication occurs from User 1 to User 2 for $n_{q-1}$ slots of time followed by communication from User 2 to User 1 for $n_q$ slots of time. There are functions(encoders) $f_1, f_2, \ldots, f_q$; if $i$ is odd, $f_i$ is used for encoding all available knowledge at User 1 into the input for channel $\mathcal C_1$ and if $i$ is odd, $f_i$ is used for encoding all available knowledge at User 2 into the input for channel $\mathcal C_2$.

Denote $x_1^n$ by $\vec{x}_1$ and $x_2^n$ by $\vec{x}_2$.

$\vec{u}_1= f_1(\vec{x}_1)$, denote $\vec{v}_1 = \mathcal C_1(\vec{u}_1)$, $\vec{u}_1 \in \mathbb I_1^{n_1}$;  $\vec{u}_2= f_2(\vec{x}_2, \vec{v}_1)$, denote $\vec{v}_2 = \mathcal C_2(\vec{u}_2)$,  $\vec{u}_2 \in \mathbb I_2^{n_2}$.

$\vec{u}_3= f_1(\vec{x}_1, \vec{u}_1, \vec{v}_2)$, denote $\vec{v}_3 = \mathcal C_1(\vec{u}_3)$,  $\vec{u}_3 \in \mathbb I_1^{n_3}$; $\vec{u}_4= f_2(\vec{x}_2, \vec{v}_1, \vec{u}_2, \vec{v}_3)$, denote $\vec{v}_4 = \mathcal C_1(\vec{u}_4)$,  $\vec{u}_4 \in \mathbb I_2^{n_4}$.

In general, for $k$ even, $k \leq q$,

$\vec{u}_{k-1}= f_1(\vec{x}_1, \vec{u}_1, \vec{v}_2, \vec{u}_3, \vec{v}_4, \ldots, \vec{u}_{k-1}, \vec{v}_{k-2})$, denote $\vec{v}_{k-1} = \mathcal C_1(\vec{u}_{k-1})$,  $\vec{u}_{k-1} \in \mathbb I_1^{n_{k-1}}$.

$\vec{u}_k= f_2(\vec{x}_2, \vec{v}_1, \vec{u}_2, \vec{v}_3, \vec{u}_4, \ldots, \vec{u}_{k-2}, \vec{v}_{k-1})$, denote $\vec{v}_4 = \mathcal C_1(\vec{u}_k)$,  $\vec{u}_1 \in \mathbb I_2^{n_k}$.

The source reproduction functions: 

$\hat {\vec{x}}_1 = g_2^n(\vec{x}_2, \vec{v}_1, \vec{u}_2, \vec{v}_3, \vec{u}_4, \ldots, \vec{u}_{q-2}, \vec{v}_{q-1})$ and
 $\hat {\vec{x}}_2 = g_1^n(\vec{x}_1, \vec{u}_1, \vec{v}_2, \vec{u}_3, \vec{v}_4, \ldots, \vec{u}_{q-1}, \vec{v}_{q-2}, \vec{u}_{q-1}, \vec{v}_q)$.

In words, from time slot $1$ to time slot $n_1$, the information available at User 1 ($\vec{x}_1$) is mapped into an input $\vec{u}_1$ for channel $\mathcal C_1$ via a function (encoder) $f_1$. $\vec{u}_1$ is communicated over Channel $\mathcal C_1$. The output of the channel is $\vec{v}_1$ which is available at User 2. From time slot $n_1+1$ to $n_1+n_2$, the information available so far at User 2 ($\vec{x}_2, \vec{v}_1$) is mapped into an input $\vec{u}_2$ for channel $\mathcal C_2$ via a function  $f_2$. $\vec{u}_2$ is communicated over the channel $\mathcal C_2$. The output of the channel is $\vec{v}_2$ which is available at User 1. From time slot $n_1+n_2+1$ to $n_1+n_2+n_3$, the information available at User 1 so far ($\vec{x}_1, \vec{u}_1, \vec{v}_2$) is mapped into an input $\vec{u}_3$ for channel $\mathcal C_1$. The output of the channel is $\vec{v}_3$ which is available at User 2. From time slot $n_1+n_2+n_3+1$ to $n_1+n_2+n_3+n_4$, the information available so far at User 2 ($\vec{x}_2, \vec{v}_1, \vec{u}_2, \vec{v}_3$) is mapped into an input $\vec{u}_4$ for channel $\mathcal C_2$ via a function  $f_4$. $\vec{u}_4$ is communicated over the channel $\mathcal C_2$. The output of the channel is $\vec{v}_4$ which is available at User 1. And so on. At the end of $q$ rounds of communication, an estimate of $\vec{x}_1$ is made at User 2 via a function (decoder) $g_2$ based on all available knowledge at User 2. This estimate is denoted by $\hat{\vec{x}}_1$.  Similarly, an estimate of $\vec{x}_2$ is made at User 1 via a function (decoder) $g_1$ based on all available knowledge at User 1. This estimate is denoted by $\hat{\vec{x}}_2$.

Denote $X_1^n, X_1^n, \hat{X}_1^n, \hat{X}_2^n$ by $\vec{X}_1, \vec{X}_2, \hat{\vec{X}}_1, \hat{\vec{X}}_1$ respectively.

The criteria for the code to be a $(\mathcal C_1, \mathcal C_2, X_1, X_2, R_1, R_2, D_1, D_2)$ code are:  $\frac{n_1 + n_3 + \cdots + n_{q-1}}{n}  \leq R_1$,  $\frac{n_2 + n_4 + \cdots + n_q}{n} \leq R_2$, $\frac{1}{n} Ed_1(\vec{X}_1, \hat{\vec{X}}_1) \leq D_1$, $\frac{1}{n} Ed_2(\vec{X}_2, \hat{\vec{X}}_2) \leq D_2$.



















\section{Source-channel separation} \label{SCS}







First, note that if there exists a block-length $n$ $(\mathcal C_1, \mathcal C_2, X_1, X_2, R_1, R_2, D_1, D_2)$ code for 2-way interactive communication with $c_{11} = 1$ and $c_{2N} = 1$, then there exists a staggered block-length $n$ $(\mathcal C_1, \mathcal C_2, X_1, X_2, R_1, R_2, D_1, D_2)$ code for $2$-way interactive communication. The basic idea of the proof is that if there is a communication from both User 1 to User 2 and  User 2 to User 1 in a certain time slot,  carry out only the communication from User 1 to User 2 in that time slot. Carry out the communication from User 2 to User 1 in the next time slot in the same manner (via the same function) as  the original code. And during this time slot when communication is happening from User 2 to User 1, do not communicate from User 1 to User 2. Then, repeat the process for the next time slot. Thus, corresponding to communication in a certain time slot in the original code, there is communication in either 1 or 2 time slots in the constructed straggered code. Clearly, this staggered code has the same $R_1, R_2, D_1, D_2$ as the original code. The time horizon of this straggered code is at most twice the time horizon of the original code but this is immaterial by definition.

If either $c_{11}$ or $c_{2N}$ is  zero in the original code (call the code  $K$) one can construct a new code with  $c_{11} = 1$ and $c_{2N} = 1$  by doing spurious communication from User 1 to User 2, use the original code from time $1$ to $N+1$ and then, during time slot $N+2$, do spurious communication from User 2 to User 1. This code, $K'$ has the same $D_1$ and $D_2$ as the original code but $R_1$ and $R_2$ are larger. In order to circumvent this problem, consider the repetition code corresponding to $K$. Thus, let the block length be $nH$, where $H$ is a natural number. Use the code $K$ from time $1$ to $N$ to communicate  $(X_1, \ldots, X_n)$ and $(X'_1, X'_2, \ldots, X'_n)$. Again use the code $K$ from time $N+1$ to time $2N$ to communicate $(X_{n+1}, \ldots, X_{2n})$ and $(X'_{n+1}, \ldots, X'_{2n})$. denote this code by $K_H$. The code $K_H$ has the same $R_1, R_2, D_1, D_2$ as the code $K$, just that the block-length is $nH$ and the time horizon is $NH$. Now, for the code $K'_H$ from $K_H$ in the way that the code $K'$ is formed from $K$. As $H$ increases, $R_1$ and $R_2$ for $K'_H$ approach those for $K_H$. Thus, given a block-length $n$ code $(\mathcal C_1, \mathcal C_2, X_1, X_2, R_1, R_2, D_1, D_2$, a block-length $nH$ staggered code$ (\mathcal C_1, \mathcal C_2, X_1, X_2, R_1 + \delta, R_2 + \delta, D_1, D_2)$ can be constructed for any $\delta > 0$.

The block-length $n$ is immaterial by definition. All that matters is that there exists a staggered code corresponding to another code with the same specifications as the original code. We have shown a construction for a staggered code  with specifications as close to the original code as required.

Without loss of generality, then, assume that the code $K$ is staggered, that the first round of communication happens from User 1 to User 2 and that, the second round of communication happens from User 2 to User 1. With the notation developed previously, let the random vectors corresponding to vectors $\vec{u}_i$, $\vec{v}_i$ be denoted by $\vec{U}_i$, $\vec{V}_i$ respectively.

Then, 
\begin{align}
                & n_1C_1 \geq^(*^0) I(\vec{U}_1, \vec{V}_1) \geq^{(*^1)}  I(\vec{X}_1, \vec{V}_1) =^{(*^2)} I(\vec{X}_1, \vec{V}_1) + I(\vec{X}_2; \vec{V}_1|\vec{X}_1) + I(\vec{X}_1; \vec{V}_2|\vec{X}_2, \vec{V}_1)
\end{align}
$(*^0)$ follows by basic properties of entropy and mutual information, $(*^1)$ above follows by data processing and $(*^2)$ follows because $\vec{V}_1-X_1-X_2$ and $X_1-(X_2,\vec{V}_1)-\vec{V}_2$ are Markoff chains.

To prove: 
\begin{align}\label{Communication1}
I(\vec{X}_1, \vec{V}_1) + I(\vec{X}_2; \vec{V}_1|\vec{X}_1) + I(\vec{X}_1; \vec{V}_2|\vec{X}_2, \vec{V}_1) \geq I(\vec{X}_1; \vec{V}_1, \vec{V}_2|\vec{X}_2) 
\end{align}
Proof:
\begin{align} 
   & I(\vec{X}_1, \vec{V}_1) + I(\vec{X}_2; \vec{V}_1|\vec{X}_1) + I(\vec{X}_1; \vec{V}_2|\vec{X}_2, \vec{V}_1) - I(\vec{X}_1; \vec{V}_1, \vec{V}_2|\vec{X}_2) \nonumber \\
=^{(*^4)} & I(\vec{X}_1, \vec{V}_1) + I(\vec{X}_2; \vec{V}_1|\vec{X}_1) + I(\vec{X}_1; \vec{V}_2|\vec{X}_2, \vec{V}_1) - [I(\vec{X}_1;\vec{V}_1|\vec{X}_2) + I(\vec{X}_1;\vec{V}_2|\vec{X}_2;\vec{V}_1)] \nonumber \\
= & I(\vec{X}_1, \vec{V}_1) + I(\vec{X}_2; \vec{V}_1|\vec{X}_1) - I(\vec{X}_1;\vec{V}_1|\vec{X}_2) = I((\vec{X}_1, \vec{X}_2; \vec{V}_1) - I(\vec{X}_1; \vec{V}_1|\vec{X}_2) \nonumber \\
= & I(\vec{X}_2; \vec{V}_1)+I(\vec{X}_1;\vec{V}_1|\vec{X}_2)-I(\vec{X}_1;\vec{V}_1|\vec{X}_2) = I(\vec{X}_2;\vec{V}_1) \geq 0
\end{align}
$(*^4)$ follows by chain rule for mutual information. Hence, proved.

In what follows, basic properties of mutual information, Data processing inequality, chain rule for mutual information and that $I(A;B|C) = 0$ if $A-C-B$ is a Markoff chain are repeatedly used and will not be pointed to each time.

Thus, 
\begin{align}
n_1C_1 \geq I(\vec{X}_1; \vec{V}_1, \vec{V}_2|\vec{X}_2)
\end{align}

Next,
\begin{align}
         & n_2C_2
\geq  I(\vec{{U}}_2; \vec{V}_2)
\geq I(\vec{X}_2, \vec{V}_1; \vec{V}_2)
=        I(\vec{X}_2, \vec{V}_1; \vec{V}_2) + I(\vec{X}_1; \vec{V}_2|\vec{X}_2, \vec{V}_1) + I(\vec{X}_2; \vec{V}_1|\vec{X}_1)
\end{align}
To prove:
\begin{align} \label{n2C2}
I(\vec{X}_2, \vec{V}_1; \vec{V}_2) + I(\vec{X}_1; \vec{V}_2|\vec{X}_2, \vec{V}_1) + I(\vec{X}_2; \vec{V}_1|\vec{X}_1) \geq I(\vec{X}_2; \vec{V}_1, \vec{V}_2|\vec{X}_1)
\end{align}
Proof: 
\begin{align}
        & I(\vec{X}_2, \vec{V}_1; \vec{V}_2) + I(\vec{X}_1; \vec{V}_2|\vec{X}_2, \vec{V}_1) + I(\vec{X}_2; \vec{V}_1|\vec{X}_1) - I(\vec{X}_2; \vec{V}_1, \vec{V}_2|\vec{X}_1) \nonumber \\
=      &  I(\vec{X}_2, \vec{V}_1; \vec{V}_2) + I(\vec{X}_1; \vec{V}_2|\vec{X}_2, \vec{V}_1) + I(\vec{X}_2; \vec{V}_1|\vec{X}_1) - [I(\vec{X}_2; \vec{V}_1|\vec{X}_1) + I(\vec{X}_2; \vec{V}_2|\vec{X}_1, \vec{V}_1)] \nonumber \\
=      & I(\vec{X}_2, \vec{V}_1; \vec{V}_2) + I(\vec{X}_1; \vec{V}_2|\vec{X}_2, \vec{V}_1) - I(\vec{X}_2; \vec{V}_2|\vec{X}_1, \vec{V}_1) \nonumber \\
=      & H(\vec{V}_2) - H(\vec{V}_2|\vec{X}_2, \vec{V}_1) + H(\vec{V}_2|\vec{X}_2, \vec{V}_1) - H(\vec{V}_2|\vec{X}_1, \vec{X}_2, \vec{V}_1) - [H(\vec{V}_2|\vec{V}_1, \vec{X}_1) - H(\vec{V}_2|\vec{X}_1, \vec{X}_2, \vec{V}_1)] \nonumber \\
=      & H(\vec{V}_2) - H(\vec{V}_2|\vec{V}_1, \vec{X}_1)
=       I(\vec{V}_2; \vec{X}_1, \vec{V}_1)
\geq  0
\end{align}
Hence proved.

Thus,
\begin{align}
n_2 C_2 \geq I(\vec{X}_2; \vec{V}_1, \vec{V}_2|\vec{X}_1)
\end{align}

Next,
\begin{align}
           n_3 C_1
\geq   I(\vec{U}_3; \vec{V}_3)
\geq   I(\vec{X}_1, \vec{U}_1, \vec{V}_2; \vec{V}_3)
\geq  I(\vec{X}_1, \vec{V}_2; \vec{V}_3)
=        I(\vec{X}_1, \vec{V}_2; \vec{V}_3) + I(\vec{X}_2, \vec{V}_1; \vec{V}_3|\vec{X}_1, \vec{V}_2) + I(\vec{X}_1, \vec{V}_2; \vec{V}_4|\vec{X}_2, \vec{V}_1, \vec{V}_3)
\end{align}

By making the transformations:
\begin{align} \label{T2to4}
\vec{V}_1  \rightarrow \vec{V}_3;
\vec{V}_2  \rightarrow \vec{V}_4;
\vec{X}_1  \rightarrow \vec{X}_1, \vec{V}_2;
\vec{X}_2  \rightarrow \vec{X}_2, \vec{V}_1
\end{align}
to (\ref{Communication1}), it follows that 
\begin{align}\label{n3c1}
I(\vec{X}_1, \vec{V}_2; \vec{V}_3) + I(\vec{X}_2, \vec{V}_1; \vec{V}_3|\vec{X}_1, \vec{V}_2) + I(\vec{X}_1, \vec{V}_2; \vec{V}_4|\vec{X}_2, \vec{V}_1, \vec{V}_3) \geq I(\vec{X}_1, \vec{V}_2; \vec{V}_3, \vec{V}_4|\vec{X}_2, \vec{V}_1)
\end{align}
That is,
\begin{align}
n_3C_1 \geq I(\vec{X}_1, \vec{V}_2; \vec{V}_3, \vec{V}_4|\vec{X}_2, \vec{V}_1)
\end{align}
Thus,
\begin{align}
(n_1 + n_3)C_1 & \geq I(\vec{X}_1; \vec{V}_1, \vec{V}_2|\vec{X}_2) + I(\vec{X}_1, \vec{V}_2; \vec{V}_3, \vec{V}_4|\vec{X}_2, \vec{V}_1) \nonumber \\
                         =  & I(\vec{X}_1; \vec{V}_1, \vec{V}_2|\vec{X}_2) + I(\vec{V}_2; \vec{V}_3, \vec{V}_4|\vec{X}_2, \vec{V}_1) + I(\vec{X}_1; \vec{V}_3, \vec{V}_4|\vec{X}_2, \vec{V}_1, \vec{V}_2) \nonumber \\
                    \geq & I(\vec{X}_1; \vec{V}_1, \vec{V}_2|\vec{X}_2) + I(\vec{X}_1; \vec{V}_3, \vec{V}_4|\vec{X}_2, \vec{V}_1, \vec{V}_2)
                         =   I(\vec{X}_1; \vec{V}_1, \vec{V}_2, \vec{V}_3, \vec{V}_4|\vec{X}_2)
\end{align}
That is, 
\begin{align}
(n_1 + n_3)C_1 \geq I(\vec{X}_1; \vec{V}_1, \vec{V}_2, \vec{V}_3, \vec{V}_4|\vec{X}_2)
\end{align}

Next,
\begin{align}
           & n_4C_2
  \geq  I(\vec{U}_4; \vec{V}_4)
  \geq  I(\vec{X}_2, \vec{V}_1, \vec{U}_2, \vec{V}_3; \vec{V}_4)
  \geq  I(\vec{X}_2, \vec{V}_1, \vec{V}_3; \vec{V}_4) \nonumber \\
  =       & I(\vec{X}_2, \vec{V}_1, \vec{V}_3; \vec{V}_4) + I(\vec{X}_2, \vec{V}_1; \vec{V}_3|\vec{X}_1, \vec{V}_2) + I(\vec{X}_1, \vec{V}_2; \vec{V}_4|\vec{X}_2, \vec{V}_1, \vec{V}_3)                 
\end{align}

By making the transformations (\ref{T2to4}) in (\ref{n2C2}), it follows that 
\begin{align}
I(\vec{X}_2, {\vec{V}}_1, \vec{V}_3; \vec{V}_4) + I(\vec{X}_2, \vec{V}_1; \vec{V}_3|\vec{X}_1, \vec{V}_2) + I(\vec{X}_1, \vec{V}_2; \vec{V}_4|\vec{X}_2, \vec{V}_1, \vec{V}_3) \geq I(\vec{X}_2, \vec{V}_1; \vec{V}_3, \vec{V}_4|\vec{X}_1, \vec{V}_2)
\end{align}
Thus,
\begin{align}
            & (n_2 + n_4)C_2 
\geq     I(\vec{X}_2; \vec{V}_1, \vec{V}_2|\vec{X}_1) + I(\vec{X}_2, \vec{V}_1; \vec{V}_3, \vec{V}_4|\vec{X}_1, \vec{V}_2) \nonumber \\
=          & I(\vec{X}_2; \vec{V}_1, \vec{V}_2|\vec{X}_1) + I(\vec{V}_1; \vec{V}_3, \vec{V}_4|\vec{X}_1, \vec{V}_2) +  I(\vec{X}_2; \vec{V}_3, \vec{V}_4|\vec{X}_1, \vec{V}_1, \vec{V}_2) \nonumber \\
\geq    & I(\vec{X}_2; \vec{V}_1, \vec{V}_2|\vec{X}_1) + I(\vec{X}_2; \vec{V}_3, \vec{V}_4|\vec{X}_1, \vec{V}_1, \vec{V}_2)
=           I(\vec{X}_2; \vec{V}_1, \vec{V}_2, \vec{V}_3, \vec{V}_4|\vec{X}_1)
\end{align}

Thus,
\begin{align}
(n_2 + n_4)C_2 \geq I(\vec{X}_2; \vec{V}_1, \vec{V}_2, \vec{V}_3, \vec{V}_4|\vec{X}_1)
\end{align}

In order to carry out induction, for some $k \geq  6$, $k$ even (note that for $k = 6$ and $k = 4$, these inequalities have already been proved above), assume that 
\begin{align}
(n_1 + n_3 + \cdots + n_{k-3})C_1 \geq I(\vec{X}_1; \vec{V}_1, \vec{V}_2, \vec{V}_3, \ldots, \vec{V}_{k-2}|\vec{X}_2) \nonumber \\
(n_2 + n_4 + \cdots + n_{k-2})C_2 \geq I(\vec{X}_2; \vec{V}_1, \vec{V}_2, \vec{V}_3, \ldots, \vec{V}_{k-2}|\vec{X}_1)
\end{align}

Then, 
\begin{align}
             n_{k-1}C_1 &
\geq      I(\vec{U}_{k-1}; \vec{V}_{k-1}) 
\geq      I(\vec{X}_1, \vec{U}_1, \vec{V}_2, \vec{U}_3, \vec{V}_4, \ldots, \vec{U}_{k-3}, \vec{V}_{k-2}; \vec{V}_{k-1})
\geq      I(\vec{X}_1, \vec{V}_2, \vec{V}_4, \ldots, \vec{V}_{k-2}; \vec{V}_{k-1}) \nonumber \\
=           & I(\vec{X}_1, \vec{V}_2, \vec{V}_4, \ldots, \vec{V}_{k-2}; \vec{V}_{k-1}) +
               I(\vec{X}_2, \vec{V}_1, \vec{V}_3, \vec{V}_{k-3}; \vec{V}_{k-1} | \vec{X}_1, \vec{V}_2, \vec{V}_4, \ldots, \vec{V}_{k-2}) + \nonumber \\
              &  I(\vec{X}_1, \vec{V}_2, \vec{V}_4, \ldots, \vec{V}_{k-2}; \vec{V}_k|\vec{X}_1, \vec{V}_1, \vec{V}_3, \ldots, \vec{V}_{k-1})
\end{align}
By making the transformations:
\begin{align} \label{TransformationsK}
\vec{V}_1 & \rightarrow \vec{V}_{k-1}; 
\vec{V}_2  \rightarrow \vec{V}_k; 
\vec{X}_1, \vec{V}_2  \rightarrow \vec{X}_1, \vec{V}_2, \vec{V}_4, \ldots, \vec{V}_{k-2}; 
\vec{X}_2, \vec{V}_1  \rightarrow \vec{X}_2, \vec{V}_1, \vec{V}_3, \ldots, \vec{V}_{k-3}
\end{align}
in (\ref{n3c1}), it follows that 
\begin{align}
              & I(\vec{X}_1, \vec{V}_2, \vec{V}_4, \ldots, \vec{V}_{k-2}; \vec{V}_{k-1}) +  
                           I(\vec{X}_2, \vec{V}_1, \vec{V}_3, \vec{V}_{k-3}; \vec{V}_{k-1} | \vec{X}_1, \vec{V}_2, \vec{V}_4, \ldots, \vec{V}_{k-2}) + \nonumber \\
                                                      &   I(\vec{X}_1, \vec{V}_2, \vec{V}_4, \ldots, \vec{V}_{k-2}; \vec{V}_k|\vec{X}_1, \vec{V}_1, \vec{V}_3, \ldots, \vec{V}_{k-1})
                                     \geq      I(\vec{X}_1, \vec{V}_2, \vec{V}_4, \ldots, \vec{V}_{k-2}; \vec{V}_{k-1}, \vec{V}_k | \vec{X}_2, \vec{V}_1, \vec{V}_3, \ldots, \vec{V}_{k-3})
\end{align}
That is, 
\begin{align}
n_{k-1}C_1 \geq I(\vec{X}_1, \vec{V}_2, \vec{V}_4, \ldots, \vec{V}_{k-2}; \vec{V}_{k-1}, \vec{V}_k | \vec{X}_2, \vec{V}_1, \vec{V}_3, \ldots, \vec{V}_{k-3})
\end{align}
Thus, by use of the induction step, it follows that
\begin{align} \label{qqqqqq}
(n_1 + n_3 + \cdots + n_{k-1})C_1 & \geq I(\vec{X}_1; \vec{V}_1, \vec{V}_2, \ldots \vec{V}_{k-2}|\vec{X}_2) + I(\vec{X}_2, \vec{V}_2, \vec{V}_4, \ldots, \vec{V}_{k-2}; \vec{V}_{k-1}, \vec{V}_k|\vec{X}_2, \vec{V}_1, \vec{V}_3, \ldots, \vec{V}_{k-3}) \nonumber \\
               &  \geq I(\vec{X}_2; \vec{V}_1, \vec{V}_2, \ldots, \vec{V}_k|\vec{X}_1)
\end{align} 
For the last inequality above, use chain rule for mutual information to decompose the second term on the left and side and ignore one of those terms, then use the chain rule for mutual information again.

\begin{align}
            n_kC_2 \geq &   I(\vec{X}_2, \vec{V}_1, \vec{U}_2, \vec{V}_3, \vec{U}_4, \ldots, \vec{U}_{k-2}, \vec{V}_{k-1}; \vec{V}_k)
\geq    I(\vec{X}_2, \vec{V}_1, \vec{V}_3, \ldots, \vec{V}_{k-1};  \vec{V}_k) \nonumber \\
                         = &      I(\vec{X}_2, \vec{V}_1, \vec{V}_3, \ldots, \vec{V}_{k-1};  \vec{V}_k)  +  
           I(\vec{X}_2, \vec{V}_1, \vec{V}_3, \ldots, \vec{V}_{k-3}; \vec{V}_{k-1} | \vec{X}_1, \vec{V}_2, \vec{V}_4, \ldots, \vec{V}_{k-2}) +  \nonumber \\
           &   I(\vec{X}_1, \vec{V}_2, \vec{V}_4, \ldots, \vec{V}_{k-1}; \vec{V}_k | \vec{X}_2, \vec{V}_1, \vec{V}_3, \ldots, \vec{V}_{k-1})
\end{align}
By use of the same set of transformations (\ref{TransformationsK}), it follows that 
\begin{align}
n_kC_2 \geq 
        I(\vec{X}_2, \vec{V}_1, \vec{V}_3, \ldots, \vec{V}_{k-3}; \vec{V}_{k-1}, \vec{V}_{k} | \vec{X}_1, \vec{V}_2, \vec{V}_4, \ldots, \vec{V}_{k-2})
\end{align}
Thus,
\begin{align}
                      (n_2 + n_4 + \cdots n_{k})C_2 
\geq              &I(\vec{X}_2; \vec{V}_1, \vec{V}_2, \ldots, \vec{V}_{k-2}|\vec{X}_1) + I(\vec{X}_2, \vec{V}_1, \vec{V}_3, \ldots, \vec{V}_{k-3}; \vec{V}_{k-1}, \vec{V}_k|\vec{X}_1, \vec{V}_2, \vec{V}_4, \ldots, \vec{V}_{k-2}) \nonumber \\
\geq  & I(\vec{X}_2; \vec{V}_1, \vec{V}_2, \ldots, \vec{V}_k|\vec{X}_1)
\end{align}
where the last step above follows in the same manner as in (\ref{qqqqqq})
It follows by induction, that
\begin{align} \label{MainMainMain}
(n_1 + n_3 + \cdots + n_{q-1})C_1 \geq I(\vec{X}_1; \vec{V}_1, \vec{V}_2, \ldots, \vec{V}_q|\vec{X}_2) \nonumber \\
(n_2 + n_4 + \cdots + n_q)C_2 \geq I(\vec{X}_2; \vec{V}_1, \vec{V}_2, \ldots, \vec{V}_q|\vec{X}_1)
\end{align} 


Assume that strict equality holds in both equations in  (\ref{MainMainMain}). Denote $\vec{V}^q \triangleq (\vec{V}_1, \vec{V}_2, \ldots, \vec{V}_q)$, $\rho_1 \triangleq I(\vec{X}_1, \vec{V}^q|\vec{X}_2)$ and $\rho_2 \triangleq  I(\vec{X}_1, \vec{V}^q|\vec{X}_2)$.  Consider Theorem 20.7 in \cite{KimGamal}, but use it to code the i.i.d. $(\vec{X}_1, \vec{X}_2)$ sources within expected distortions $nD_1$ and $nD_2$. It follows by Theorem 20.7 that rate pair $(\rho_1, \rho_2)$ is achievable for source-coding the i.i.d. $(\vec{X}_1, \vec{X}_2)$ source-pair within distortion pair $(nD_1, nD_2)$. One needs to note, for this, that the conditional pmf conditions are met and the cardinality bound conditions on the sets in Theorem 20.7 in \cite{KimGamal} are not necessary for that theorem to hold. By (\ref{MainMainMain}), rate-pair $(\rho_1, \rho_2)$ is also achievable for reliable communication from (User 1 to User 2, User 2 to User 1) by one full use of the given interactive network over the $q$ rounds of communication. It follows, then, that it is sufficient to restrict attention to source-channel separation architectures for two-way interactive communication with fidelity criteria, where by a source-channel separation architecture, we mean an architecture as follows: there exist $(z_1, z_2, \ldots, z_r)$, $r$ even for which first, reliable communication of $z_1C_1$ bits is carried over $\mathcal C_1$ from User 1 to User 2. Then, reliable communication of $z_2C_2$ bits is carried over $\mathcal C_2$ from User 2 to User 1. Then, reliable communication of $z_3C_1$ bits is carried over $\mathcal C_1$ from User 1 to User 2. Then, reliable communication of $z_4C_2$ bits is carried over $\mathcal C_2$ from User 2 to User 1. And so on. The optimality is in the sense that the number of channel uses of $\mathcal C_1$ and the number of channel uses of $\mathcal C_2$ per input source symbol  is the same in the source-channel separation scheme as in the original scheme, and the same distortion levels are achieved; this is our measure of quality of a code: as stated in Section \ref{Formulation}, the length of the time-horizon $N$ does not matter (nor does the order in which $\mathcal C_1$ and $\mathcal C_2$ are used); what matters is the number of channel uses for each block-length.

\section{Recapitulation and research directions}

Recapitulation: The channel-coding problem of 2-way interactive communication was formulated and it was proved that when the channels are memoryless, discrete, and sources are i.i.d, though possibly dependent, it is sufficient to consider source-channel separation based architectures for communication with fidelity criteria.

Research directions: Consider the case when channels are coupled. Consider generalizations to non-i.i.d. sources and non-memoryless channels.

\end{document}